\documentclass[english,twocolumn,prb]{revtex4}
\usepackage{graphicx}
\usepackage{babel}

\makeatother

\begin{document}

\title{Band engineering in graphene with superlattices of substitutional defects}
\author{Simone Casolo$^{a}$, Rocco Martinazzo$^{a,b}$, Gian Franco Tantardini$^{a,b}$}
\email{simone.casolo@unimi.it}
\email{rocco.martinazzo@unimi.it}

\affiliation{
$^{a}$Dipartimento di Chimica Fisica ed Elettrochimica,
Universit\'a degli Studi di Milano, via Golgi 19, 20133 Milan, Italy.\\
$^{b}$CIMAINA, Interdisciplinary Center of Nanostructured Materials and
Interfaces, Milan, Italy.
}

\begin{abstract}
We investigate graphene superlattices of nitrogen and boron substitutional defects and
by using symmetry arguments and electronic structure calculations
we show how such superlattices can be used to modify graphene
band structure. Specifically, depending on the superlattice symmetry,
the structures considered here can either preserve the Dirac cones ($D_{6h}$ superlattices)
or open a band gap ($D_{3h}$). Relevant band parameters (carriers effective masses,
group velocities and gaps, when present) are found to depend on the
superlattice constant $n$
as $1/n^p$ where $p$ is in the range $1-2$,
depending on the case considered. Overall, the results presented here show how one
can tune the graphene band structure to a great extent by modifying few
superlattice parameters.
\end{abstract}

\maketitle

\section{Introduction}

Single layer graphene is a very promising material for a future 
silicon-free nanoelectronics. The peculiar character of its charge 
carriers comes from the intersection of the $\pi/\pi^*$ electronic bands occurring at the corners 
of its hexagonal Brillouin zone. This gives rise to the so-called Dirac cones at the Fermi level 
and makes graphene a zero-gap semiconductor \cite{GrapheneSWM} in which  
low energy excitations behave as massless, 
chiral Dirac particles\cite{castroneto09,ChakrabortyRev}. 
In turn, this implies a series of interesting physical effects that open new
perspectives for fabricating novel electronic devices\cite{Schwierz2010},  
\emph{e.g.} high-performance transistors for radiofrequency 
applications\cite{Transistor,AvourisPerspective}. 
In this perspective the possibility of engineering 
graphene's band structure by introducing defects, strains or 
external potentials 
has gained importance in the recent past, in particular for opening a 
gap in the band structure 
which is essential to design logic devices. 
Indeed the non-vanishing residual conductance of intrinsic graphene 
avoids the complete current pinch-off in the pristine 
material\cite{PeresTransport,Avouris07}, 
thereby limiting the on-off ratio to $\sim10^1-10^2$. 
A number of controlled techniques for energy band engineering have been 
proposed other than the actively pursued goal to obtain nanoribbons 
of controlled size and edge geometry. 
Most of them are based on the use of superlattices of external potentials 
\cite{TiwariSuperlattice,ParkLouie08} or defects such as 
holes\cite{Antidot1,Liu09} 
and adsorbates \cite{Chernozatonskii}. 
Controlled vacancies on graphene \cite{Fishbein08}, 
as well as large holes symmetrically arranged to form 
graphene anti-dots \cite{GrapheneHoles} 
have actually been realized with modern lithographic and 
self-assembling techniques. 
Preferential sticking of atoms induced by Moir\'e patterns 
\cite{HornekaerNature} or by other electronic effects 
\cite{Cheianov09,Cheianov10} could also induce a superlattice ordering that 
modifies graphene energy bands. Likewise, there is a great hope that 
novel bottom-up techniques \cite{RuffieuxChemComm} 
may be applied  
to fabricate atomically precise graphenic structures 
as already shown for nanoribbons\cite{RuffieuxNature}.
These approaches might allow to realize in the near future graphene-related 
two-dimensional materials with \emph{modified} characteristics, 
\emph{e.g.} linearly-dispersing bands with variable Fermi velocities or 
semiconducting structures. 
In this paper we focus on atomically precise superlattices 
of substitutional atoms.
The present work connects to and extend a recent work\cite{Martinazzo10}
 where we have shown that 
in properly designed superlattices of holes or adatoms one can open a gap 
\emph{without} breaking graphene point symmetry, \emph{i.e.} preserving  
the pseudo-relativistic behaviour of charge carriers which makes graphene so 
attractive. The structures suggested in Ref.\citep{Martinazzo10} have  $\pi$
vacancies (hence missing $p_z$ orbitals) at the sites of a honeycomb 
superlattice, as a consequence of the 
introduction of C vacancies (holes) or chemisorption of simple atomic species. 
Here we consider similar, highly symmetric structures, but with $\pi$ 
vacancies replaced by boron and nitrogen.
Similar defects have been recently considered for tuning the electronic 
properties 
of graphene nanoribbons and other carbon based structures suggesting that, 
when arranged to form particular structures, they can turn the material into 
a semiconductor or a half-metal\cite{BNHalfMetal,PatiJPCB,ZengBN}.
Half-metallicity and the other many-body effects in such a
structures open new perspectives in the field of carbon-based materials 
for spintronic
applications: for a recent review see Ref.\cite{BN_nanoscalereview,PatiReview}
and references therein.\\
In this paper we show that, depending on the superlattice symmetry, one can 
obtain either electron (hole) doped substrates with pseudo-relativistic 
massless 
carriers or semiconducting structures with a quasi-conical dispersion, and 
with the help of electronic structure calculations (tight-binding and density-functional theory) 
we determine carriers velocities, effective masses and band gaps (when present) as functions of the superlattice . 
The focus is on boron and nitrogen, mainly because of the fast progresses 
in methods for the controlled synthesis of B and N doped graphenes.
For instance, Panchakarla \emph{et al.}\cite{Panchakarla09} have
recently shown how it is possible to insert B or N dopants in graphene by
adding the correct precursors in the arc discharge chamber, while Ci \emph{et al.}\cite{ciNature10} 
have reported the synthesis of large islands of boron
nitride embedded in graphene by atomic layer deposition techniques. Methods
to selectively replace C atoms in the graphene lattice have also been
proposed \cite{Pontes09}, thereby suggesting that the superlattice structures
considered in this paper might soon become feasible.\\
The paper is organized as follows. In the next section we summarize the 
computational details of the calculations. 
Then, we show how $p$-($n$-) doped graphene-like structures result  
when substitutional defects are arranged in \emph{honeycomb} superlattices, 
whereas  semiconducting structures with quasi-conical dispersion (massive 
Dirac carriers) result either from a \emph{hexagonal} superlattice of from 
a honeycomb co-doped superlattice. Finally, we summarize and conclude.\\
Throughout this paper we define the superlattice periodicity using Wood's notation, \emph{i.e.}
by multiplying graphene's two-dimensional lattice vectors by the integer 
(superlattice) constant $n$.

\section{Computational Methods}\label{details}

The results shown in the next sections have been obtained 
from both tight-binding (TB) 
and density functional theory (DFT) electronic structure calculations.  
In the first case we diagonalized the usual tight-binding Hamiltonian 
for graphene $\pi-\pi^*$ system, applying periodic boundary conditions 
and including hopping terms up to the third nearest-neighbors. 
The on-site energies $\epsilon_i$ and hopping terms $t_1$, $t_2$ and $t_3$ 
(for nearest, next-to-nearest and next-to-next-nearest neighbors, 
respectively) 
are those proposed by Nanda \emph{et al.}\cite{Nanda}. 
They were fitted to accurate all-electron calculations to correctly reproduce 
the Fermi velocity of single layer graphene. 
For the dopant atoms we only considered hoppings to nearest neighbor sites. 
Their values ($t_1$), as well as those of the on-site energies ($\epsilon_i$), 
are those introduced by Peres \emph{et al.}\cite{PeresEPL}, who have 
already successfully used them to study electronic effects in doped graphene. 
A summary of the TB parameters is listed in \ref{tab:TB}. \\
\begin{table}
\caption{\label{tab:TB} 
Parameters used in the tight-binding Hamiltonian. All the values are in eV.}
\centering
\vspace{0.30cm}
\begin{tabular}{@{}ccccc}
\hline\hline
Atom & $\epsilon$ & $t_1$ & $t_2$ & $t_3$ \\
\hline
C &  0.000 & -2.900 & +0.175 & -0.155 \\
B & -1.5225 & +1.450 &    -  &   -    \\
N & +1.5225 & -1.450 &    -  &   -    \\
\hline\hline
\end{tabular}
\end{table}
First principles DFT calculations were
preformed with the help of the VASP suite \cite{VASP1,VASP2}, using a 
supercell approach. Core electrons were taken into account 
by projector augmented wave (PAW) pseudo-potentials while for valence a 
500 eV plane wave cutoff was used. To correctly represent the defect 
induced charge inhomogeneities we used the Perdew-Burke-Ernzerhof 
(PBE) gradient-dependent exchange and correlation functional \cite{PBE1}.
Band structures were sampled by a $\Gamma$ centered $k$-points grid, 
never sparser than 6x6x1 in order to include every special point in the Brillouin zone (BZ).\\ 
The TB parametrization was tested by comparing the band structure of few 
superlattices along the $\Gamma$-K-M-K'-$\Gamma$ path with accurate 
DFT results. 
In every case the adopted parametrization was found to be accurate enough 
to reproduce the bands close to the Fermi energy.\\
Therefore, we computed DFT band structures 
for $n$x$n$ graphene superlattices up to $n$=14, and for larger structures 
we relied on TB calculations only.

\section{Results and Discussion}\label{results}

Graphene's peculiar electronic structure is strictly related to the point symmetry of its lattice, 
$D_{6h}$ in the Sh{\"o}nflies notation. 
In the Brillouin zone, for each Bloch electronic state 
with $k$ vector $\bf{k}$, the relevant symmetry elements are those which either leaves $\bf{k}$
invariant or transforms it into one of its equivalent images, \emph{i.e.} 
$\bf{k}\rightarrow\bf{k}+\bf{G}$ being $\bf{G}$
a reciprocal lattice vector. These elements form a subgroup of $D_{6h}$, 
known as little co-group or simply $k$-group at $\bf{k}$ \cite{Mirman}, 
which determines the possible symmetries of the electronic states at $\bf{k}$.
At the high symmetry point $K$ (or $K'$) of graphene's Brillouin zone the $k$-group
is $D_{3h}$, and Bloch functions built as linear combinations of $p_z$ orbitals
span a two-dimensional irreducible representation (irrep) of such a symmetry group ($E''$). 
This is enough for the $\pi-\pi^*$ degeneracy and the unusual linear dispersion at $K$ ($K'$).
That this occurs exactly at the Fermi level is a consequence of the electron-hole ($e-h$) symmetry
which approximately holds in graphene. Indeed, thanks to this extra symmetry, energy levels  
are always symmetrically arranged and, at half-filling, the Fermi level lies exactly at the center 
of the spectrum, where any doubly degenerate level is forced to lay \footnote{ Notice that even though $e-h$ symmetry only holds in the nearest neighbors approximation and 
in absence of diagonal disorder, the Fermi level always matches the doubly degenerate state at $K$ ($K'$)   
as long as the $e-h$ symmetry breaking does not cause the maximum (minimum) of the valence (conduction) band 
to exceed the energy at $K$. }. In general, the number of doubly degenerate irreps in the BZ determines alone the presence 
of states (absence of a gap) at the Fermi level. We have recently shown\cite{Martinazzo10} how one can turn such number 
to be even at every special point -thus opening a gap in the band structure- by symmetrically removing ``$p_z$ 
orbitals'' in forming certain $n$x$n$ superlattices.  
Substitutional defects behave similarly to $p_z$ vacancies (to which they reduce when the hoppings become zero) 
but introduce impurity bands which partially hybridize with those of the substrate. 
In addition, the diagonal disorder they introduce breaks $e-h$ symmetry giving rise to a Fermi level shift, 
$i.e.$ to $p-$ and $n-$ doping for group IIIA and VA elements, respectively, as recently shown for both graphene \cite{Roche08} and nanotubes \cite{Zheng2010}
. In the weakly defective superstructures considered in the following  
the defect-induced perturbation affects the electronic structure close to the Fermi level.
With homogeneous doping the latter shifts at most proportionally to $1/n$, 
\emph{i.e.} as the square root of the defect concentration, 
as a consequence of the linear-energy dispersion  which implies $E_F=v_F\sqrt{\pi n_e }$, 
where $v_F$ is the Fermi velocity of pristine graphene and 
$n_e$ is the electron (hole) excess density, $n_e\propto1/n^2$. 
Hence, analogously to the superlattices of $p_z$ vacancies\cite{Martinazzo10}, 
we make use of symmetry arguments to establish whether degeneracy occurs at the BZ special points 
in the important low-energy region.
\begin{figure}
\begin{centering}
\includegraphics[clip,width=0.88\columnwidth]{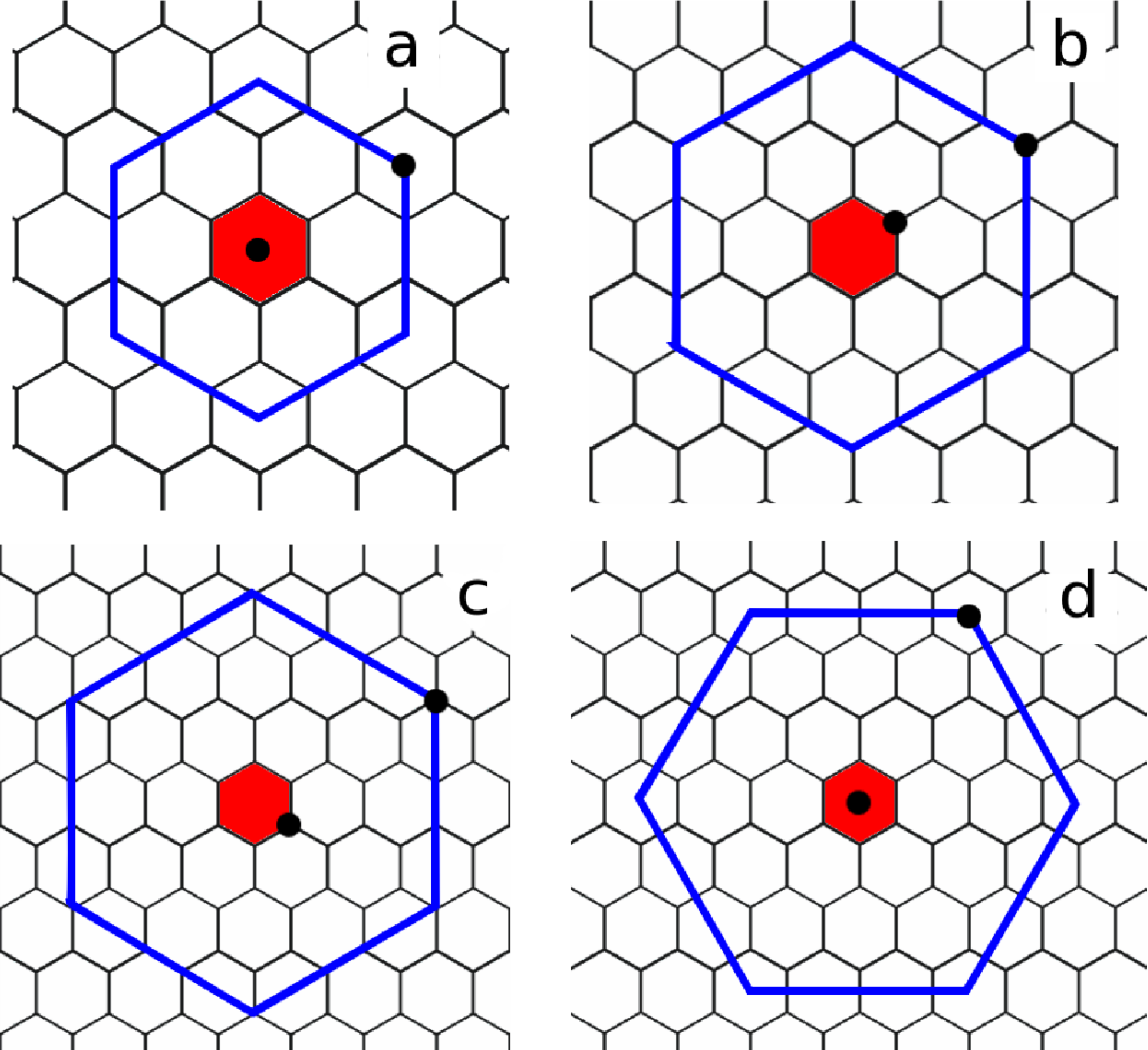}
\par\end{centering}
\caption{\label{figfold}Folding of graphene Brillouin zone (BZ, blue line) 
into the superlattice ones (red filled hexagon) for some $n$x$n$ structures, a) $n$=3$m$, b) $n$=3$m+1$ and c) $n$=3$m$+2, along with 
the case of $\sqrt{3}n$x$\sqrt{3}nR30^{\circ}$ superlattices (d). 
The $K$ point in graphene's BZ, is labeled with a black filled 
dot.}
\end{figure}
It is worth noticing at this point that, however small the defect perturbation is, 
the folding of graphene band structure occurs differently according 
whether the superlattice constant $n$ belongs to the sequence $n=3m$ or $n=3m+1,3m+2$ ($m$ integer). 
As shown in \ref{figfold} for the superlattices considered in this work (a-c), 
for $n=3m+1$ ($3m+2$) $K$ and $K'$ fold separately into $K_n$($K'_n$) and 
$K'_n$($K_n$), whereas for $n=3m$ they both fold to the BZ center $\Gamma_n$. 
This means that $n=3m$ superlattices are expected to have rather unique properties 
related to the highly degenerate nature of the unperturbed spectrum. 
In the following we mainly focus on $n=3m+1,3m+2$ superlattices and 
only occasionally look at the properties of $n=3m$ ones. A further six-fold superlattice symmetry, 
the  $\sqrt{3}n$x$\sqrt{3}nR30^{\circ}$ case reported in \ref{figfold}(d), 
will not be considered here since in that case band folding occurs analogously to the $3m$x$3m$ case discussed above.

\subsection{Honeycomb superlattices}\label{honeycombs}

\begin{figure}
\begin{centering}
\includegraphics[clip,width=0.89\columnwidth]{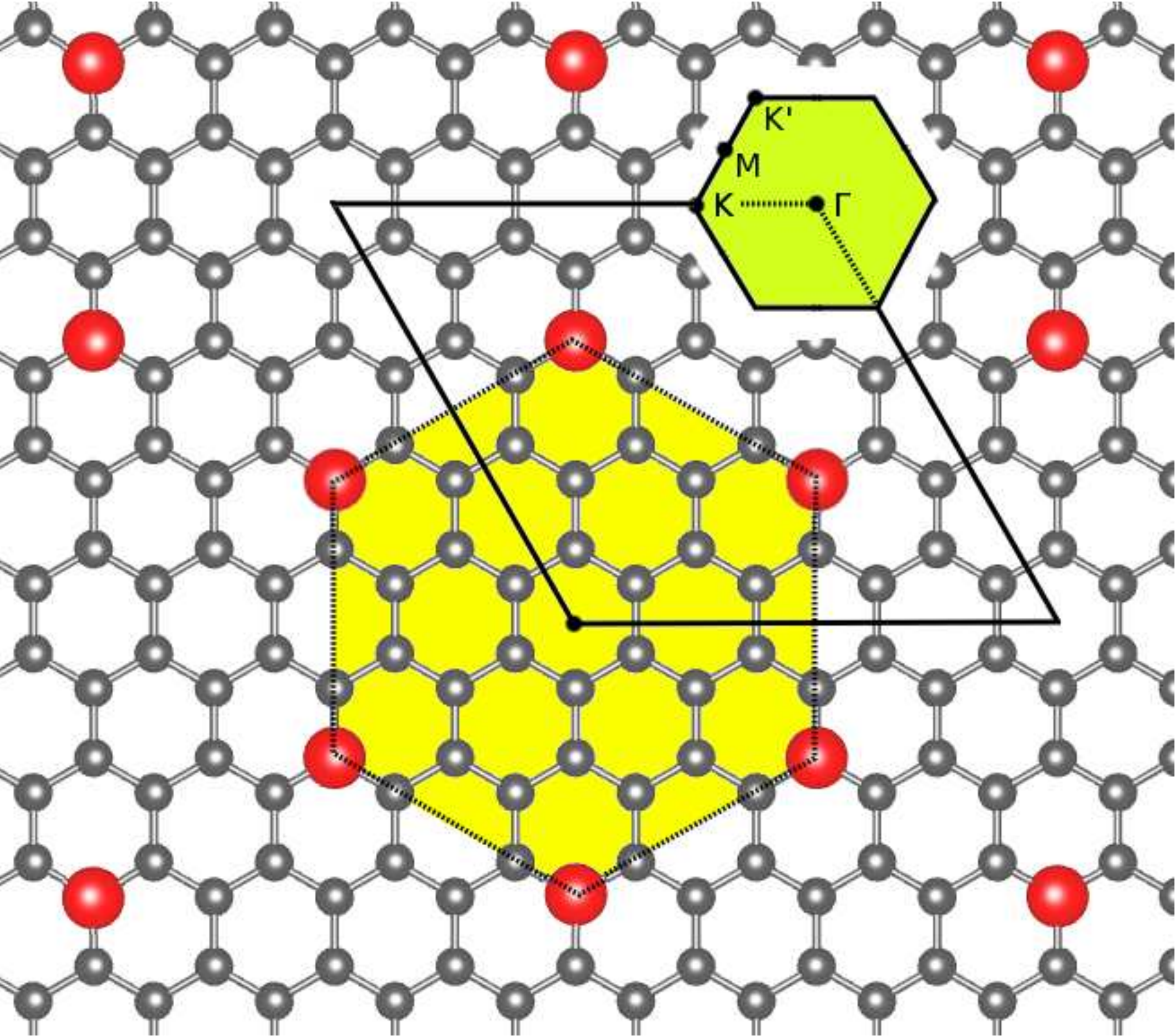}
\par\end{centering}
\caption{\label{honey}A 4x4-honeycomb superlattice: the black line represents the unit cell 
while the Wigner-Seitz and Brillouin zone are shown in yellow and green 
respectively. Red balls are sublattice substitutional defects forming the 
superlattice.}
\end{figure}

A honeycomb-shaped superlattice is a natural choice for $n$x$n$ 
superlattices ($n$x$n$-honeycombs thereafter), 
since it preserves the $D_{6h}$ point group symmetry of pristine graphene.
The superlattice unit cell contains two substitutional atoms and is shown in \ref{honey}. 
If the atomic radii of the dopants are small enough that lattice distortions  are minimal, 
the system overall symmetry is preserved and Dirac cones at $K_n$ and $K'_n$ are expected. 
This is the case of boron and nitrogen substitutional defects, whose DFT-optimized structures 
show no appreciable lattice distortion. Both TB and DFT calculations confirm that $n=3m+1$ and $3m+2$ 
honeycomb superlattices made of B or N substitutional defects only show a low-energy band structure very 
similar to that of perfect graphene, but with the Fermi level lying respectively below ($p$-doped) 
and above ($n$-doped) their Dirac point. In principle, with properly designed $n-$ or $p-$ back-doping, 
\emph{e.g.} electric-field induced but also \emph{via} molecular adsorption\cite{moldoping10,moldoping10bis}, 
such shift can be offset and the analogy with pristine graphene can be fully exploited.\\ 
\begin{figure}
\begin{centering}
\includegraphics[clip,width=0.89\columnwidth]{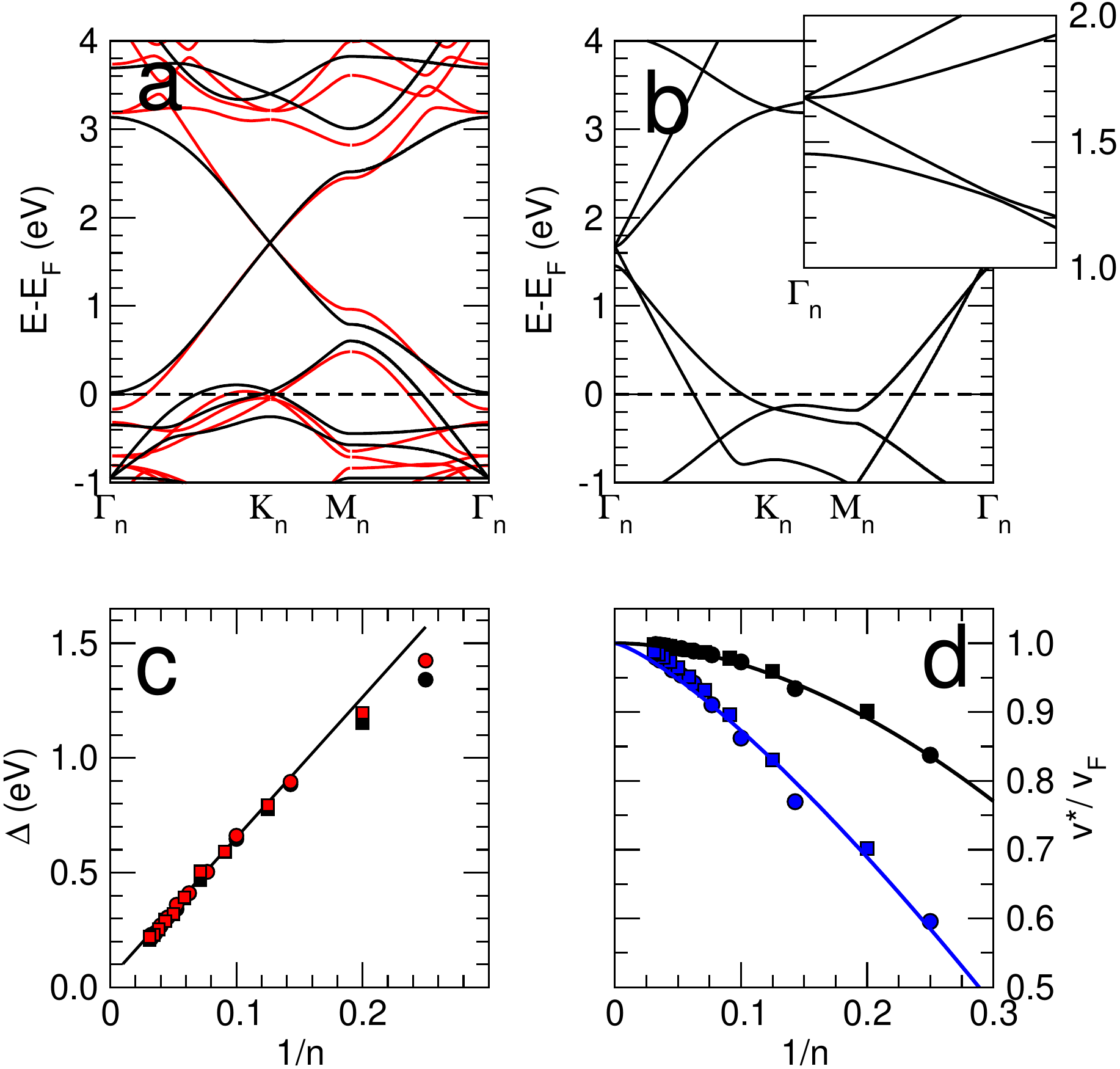}
\par\end{centering}
\caption{\label{fighoneybands} $a$): TB (black lines) and DFT (red lines) band structures
for the 4x4-honeycomb boron superlattice. $b$) the TB band structure of the 3x3-honeycomb 
boron superlattice arising from folding in $n=3m$ superlattices. 
The inset shows a close-up of the region close to $\Gamma_n$. $c$) Absolute shift of the 
Dirac cones apex ($\Delta$) with respect to the Fermi level, in $p$-doped (B, red) and, 
$n$-doped (N, black) honeycombs. $d$) Group velocity for charge carriers close to the cones apex for the 
boron (black) and nitrogen (blue) case. Circles and squares for $n=3m+1$ and $n=3m+2$.}
\end{figure}
%
\ref{fighoneybands}(a) shows the TB and \emph{first principles} 
band structures of one $n$x$n$-honeycomb together with the position of 
the Fermi level (\ref{fighoneybands}(c))
in such $n$- and $p$-doped superlattices at different impurity concentrations. 
As expected, the shift ($\Delta$) of the Dirac cones with respect to the Fermi level (see \ref{fighoneybands}(c)) 
is, to a good approximation, inversely proportional to the dopant concentration for both B and N doping, 
though with opposite sign. 
The difference between TB and DFT band structure is minimal, and 
this confirms that the tight binding parameters adopted are good enough for accurately 
describing the low-energy features of the $n$x$n$-honeycomb superlattices investigated in this paper.  
In \ref{fighoneybands}(b) we also report the unique band structure resulting from the special folding in the $n=3m$ sequence; 
as it is evident from the inset of \ref{fighoneybands}(b), the four-fold degeneracy occurring at $\Gamma_n$ is partially lifted, 
and a gap is introduced in one of the two cone replica.

The group velocity for electrons and holes taken close 
to the cone apex (but rather adequate for a wider energy range) is shown in \ref{fighoneybands}(d)  
for $n$- and $p$-doped superlattices. The two curves approach the limit of clean graphene 
with different trends. Upon non-linear curve fitting the group velocity $v$ 
(relative to the one in pristine graphene) for $p$-doped honeycombs is found to behave 
as $v/v_F\propto 1-n^{-1.29}$, while for $n$-doped honeycombs as $\propto 1-n^{-1.84}$.
The difference between the two cases is due to the value of the on-site 
energies and hopping of the dopants which determine the degree of hybridization 
of their impurity levels with that of bulk graphene. With the parameters used (see \ref{tab:TB}), 
which are symmetric with respect to the on-site energy of C atoms, this can only happen because 
of the asymmetry in graphene electronic structure introduced by the 
next-to-nearest neighbor interactions.
Other superlattices made of group IIIA (Al, Ga, In)  
and VA (P, As, Sb) dopants have been tested by \emph{first principles} 
calculations. In any case we found that, after geometric optimization of the lattice 
structure, the impurities stand out from the graphene layer plane and 
considerably distort the neighboring lattice positions. 
The resulting band structures are metallic but lack of Dirac cones 
due to the reduced symmetry.

\subsection{Hexagonal superlattices}\label{hexagonal}

\begin{figure}[h]
\begin{centering}
\includegraphics[clip,width=0.80\columnwidth]{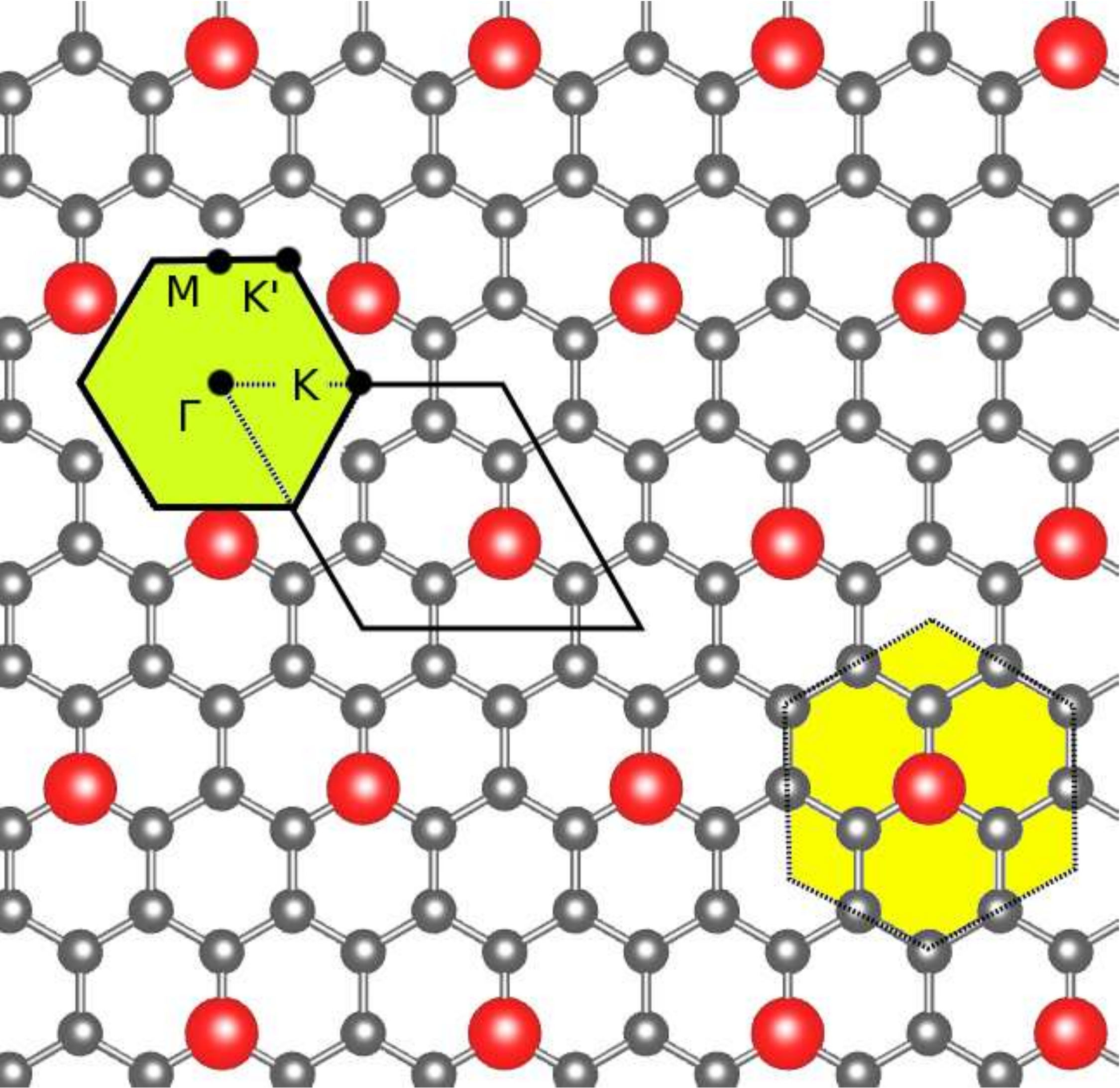}
\par\end{centering}
\caption{\label{fig:hex}
A 2x2-hexagon superlattice: the black line represents the unit cell boundary
while the Wigner-Seitz and Brillouin zone are filled in yellow and green 
respectively. Red balls are substitutional defects positions.
}
\end{figure}
\begin{figure}
\begin{centering}
\includegraphics[clip,width=0.80\columnwidth]{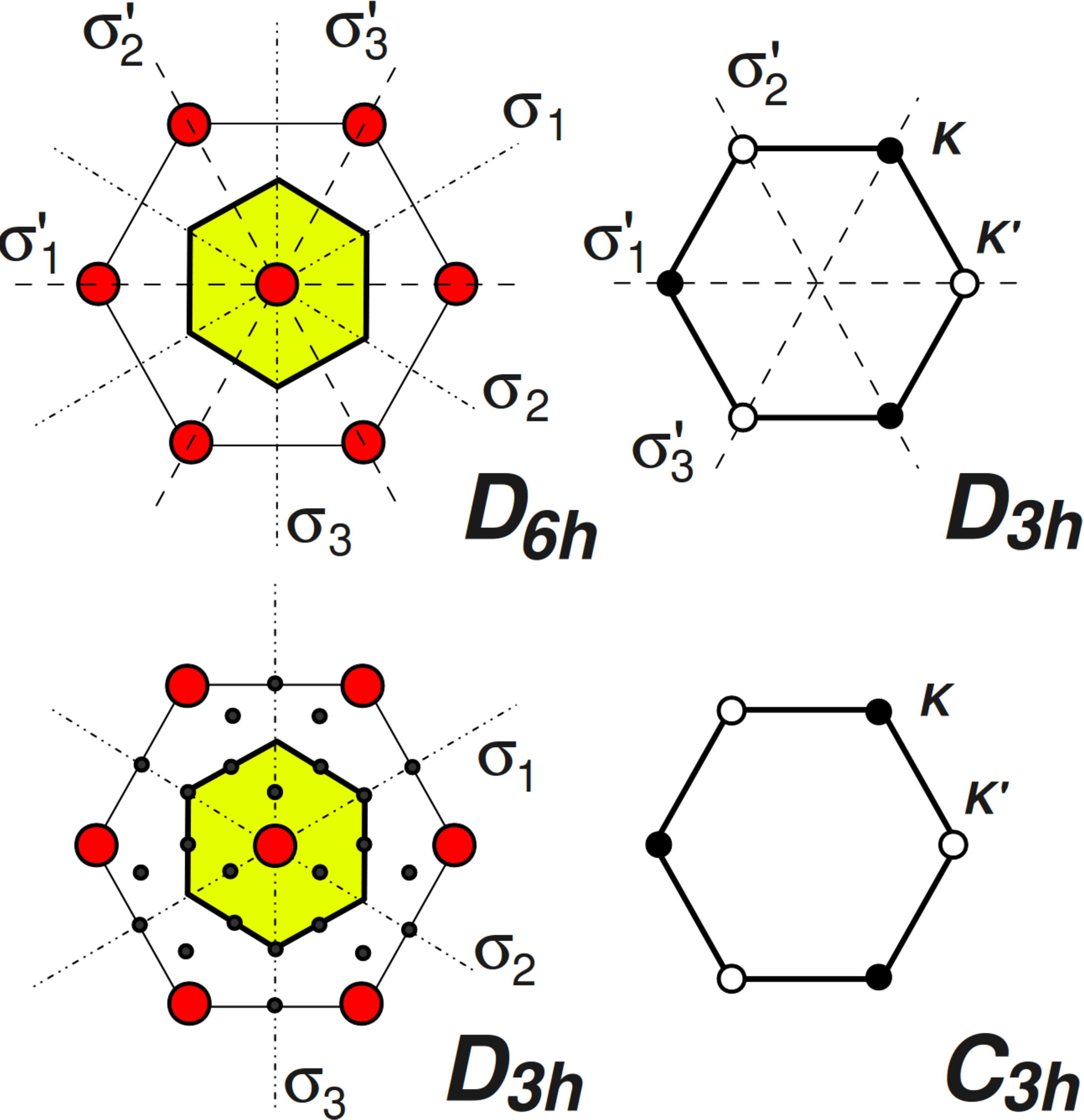}
\par\end{centering}
\caption{\label{D6hD3hHexa}
Wigner-Seitz (yellow, left side) and Brillouin zone (right side)
for the hexagonal superlattices. Considering the defects
only the point group is $D_{6h}$, and the corresponding $k$-group in $K_n$ 
$D_{3h}$ (upper panel). Overall the underlying carbons remove the $\sigma '$
planes, reducing the symmetry to $D_{3h}$.
}
\end{figure}
\begin{figure}
\begin{centering}
\includegraphics[clip,width=1.1\columnwidth]{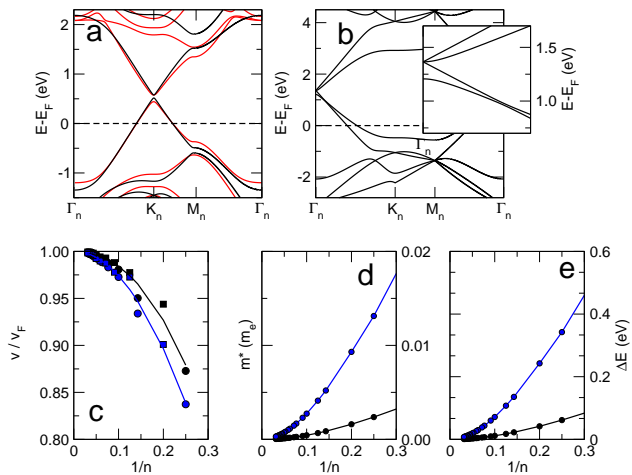}
\par\end{centering}
\caption{\label{hexa}$a$) TB (black) and DFT (red) band structures for the 4x4-hexagon. 
$b$) the TB band structure of the 3x3-hexagon boron superlattice arising from folding 
in $n=3m$ superlattices. The inset shows a close-up of the region close to $\Gamma_n$.
Carriers effective speeds ($c$), masses ($d$) and energy band-gaps ($e$) versus $1/n$. 
Black and blue values refer to B and N superlattices, respectively.}
\end{figure}

When one defect per supercell only is introduced a $n$x$n$ 
hexagonal superlattice 
(a ``$n$x$n$-hexagon'') results, as shown in \ref{fig:hex}. 
This kind of structures is closely related to the honeycomb ones, 
having one extra substitutional atom at the center of a hexagon 
of defects. A closer inspection, however, reveals that, due to the presence of 
the underlying C network, the point symmetry is reduced to $D_{3h}$, 
with $\sigma$ planes missing with respect to the honeycomb counterparts. 
It follows that the $k$-group at $K_n$ ($K'_n$) is $C_{3h}$, 
with no irreducible two-dimensional (complex) representations (see \ref{D6hD3hHexa}). 
Hence, degeneracy is removed at the special points and a (small) gap opens in the band 
structures, close to the (shifted) Fermi energy. This is shown in \ref{hexa}(a) 
where the TB and DFT band-structures of the $4$x$4$ hexagon are reported. 
The energy spectrum of such gapped graphene is compatible with charge carriers
behaving as \emph{massive} Dirac particles 
\begin{equation}\label{eq:relE}
E({\bf{k}})=\pm v\sqrt{k_x^2+k_y^2+m^{*2}v^2}
\end{equation} 
where $v,m^*$ are the effective `speed of light' and `rest mass', respectively, and determine the gap size
\begin{equation}\label{eq:mass}
\Delta E = 2 m^* v^2
\end{equation}
According to the semiclassical theory of conduction $m^*$ is also the effective mass $m_{eff}$ governing charge carrier 
mobility for $k<<mv$; for $k>>mv$ carriers behave pseudorelativistically with $m_{eff}=0$ and limiting speed $v$.  
The values $v$, $m^*$ and $\Delta E$ have been obtained by non-linear curve fitting 
of the numerical results to \ref{eq:relE} and are reported in panels (b)-(d) of \ref{hexa}.
For $n$x$n$-hexagons the band gap is very dependent on the type of 
dopant (\ref{hexa}.(e)): the maximum gaps, occurring in $2$x$2$ hexagons,
are 0.93 eV for nitrogen and only 0.17 eV for boron. 
The effective masses of electrons and holes (\ref{hexa}.(d)) 
roughly scale as the gaps: $\propto n^{-1.45}$ and $\propto n^{-1.52}$ 
for $n$- and $p$-doped structures respectively, and their maximum is 3.7x$10^{-2}$ and 6.7x$10^{-3}$ $m_e$. 
This is similar to the case of graphene nanoribbon\cite{ScuseriaRibbons}, 
whose band gap scales as the inverse of their width even though here
the gap is due to symmetry breaking rather than quantum confinement.
The shift of the Fermi level (not shown) is again proportional 
to the square root of the defects concentration, that is now only  half of the value 
for honeycombs with the same superlattice periodicity. 
Charge carriers velocities scale similarly for the two dopant species 
as shown in \ref{hexa}.(c) with a best-fit exponent close to -2 
($v(B)/v_F\propto 1-n^{-1.98}$, $v(N)/v_F\propto 1-n^{-2.28}$).
In \ref{hexa}, panel (b), we also report the particular band structure arising in $n=3m$ hexagon superlattices. 
At the relevant special point $\Gamma_n$, the massive, pseudo-relativistic energy dispersion 
is superimposed with a massless one, thereby giving rise to a two-valley system with very different charge carriers.  
As shown in the next subsection, all the features discussed in this Section can be brought 
at the Fermi level by co-doping the substrate in forming a honeycomb structure with the same $D_{3h}$ 
symmetry discussed here.

\subsection{Co-doped superlattices}\label{codoped}

\begin{figure}
\begin{centering}
\includegraphics[clip,width=0.80\columnwidth]{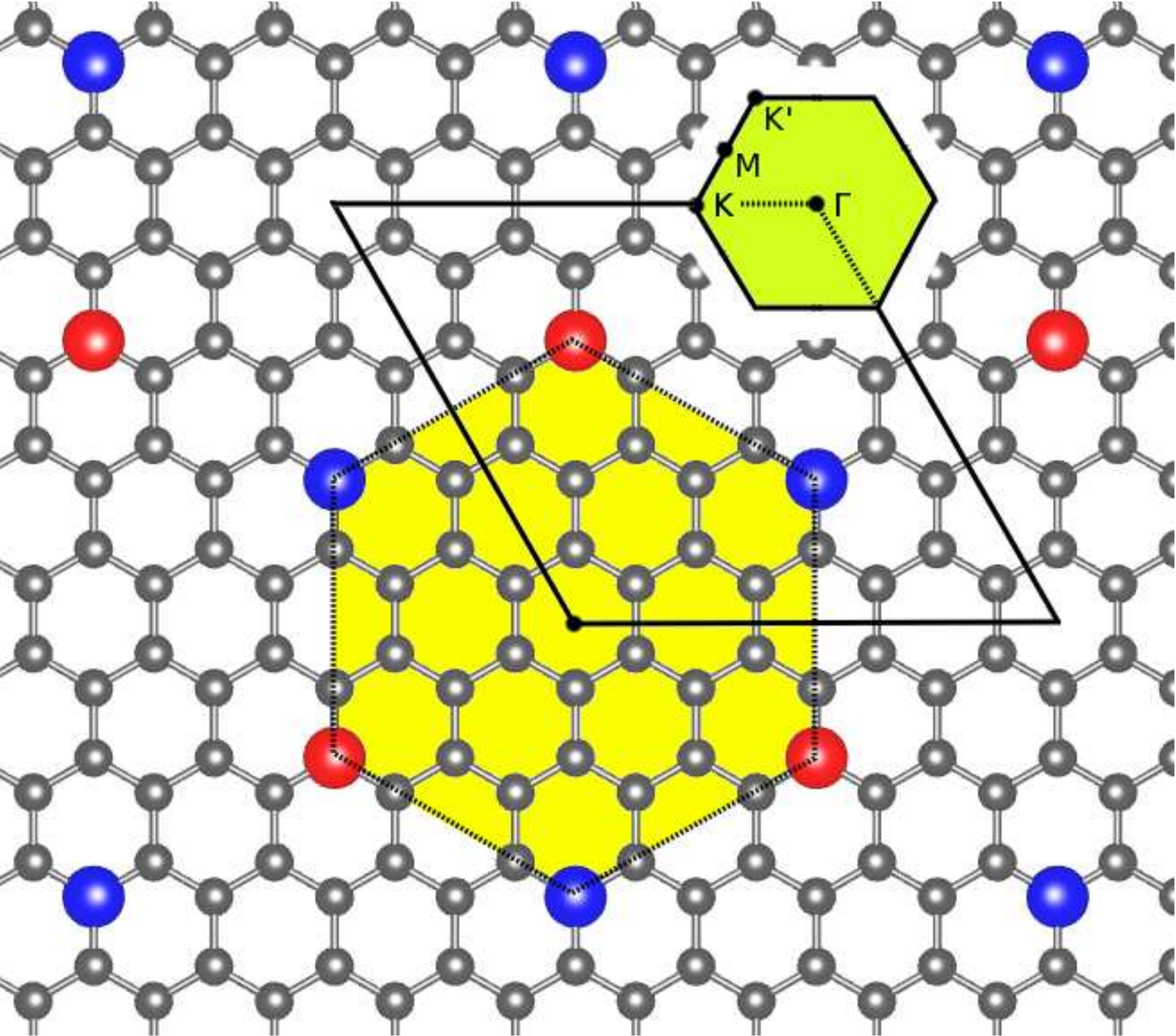}
\par\end{centering}
\caption{\label{cohoney}A 4x4-honeycomb superlattice: 
the black line represents the unit cell while the Wigner-Seitz and Brillouin zone 
are shown in yellow and green respectively. 
Red and Blue balls are B and N substitutional defects.}
\end{figure}

One further possible superlattice arrangement is obtained
by using two different dopants in the $n$x$n$-honeycomb unit cell, \emph{i.e.}
co-doping the structures with boron and nitrogen (see \ref{cohoney}). 
In this way  B and N atoms form a boron nitride-like honeycomb superlattice in which 
sublattices equivalence (and symmetry) is broken.
This is analogous to place graphene in the modulating field of a proper substrate, 
\emph{e.g} a hexagonal BN (0001) surface, which has been shown to lift the degeneracy of the
 $\pi-\pi^*$ bands \cite{Giovannetti07}; similarly for deposition, or growth, on silicon carbide surfaces 
\cite{Mattausch07,VarchonBerger}. The superlattice structures considered here
offer the possibility to modify the periodicity of the perturbation, 
and thus to tune the gap. Indeed, this kind of superlattices present $D_{3h}$ point symmetry, 
hence a $C_{3h}$ $k$-group in $K_n, K'_n$, and,  analogously to the hexagonal case discussed above, 
open a band gap typical of massive Dirac particles. Differently from before, however, 
the structures considered here are iso-electronic with graphene and therefore  
the gap lies exactly at the Fermi energy.
\begin{figure}
\begin{centering}
\includegraphics[clip,width=0.99\columnwidth]{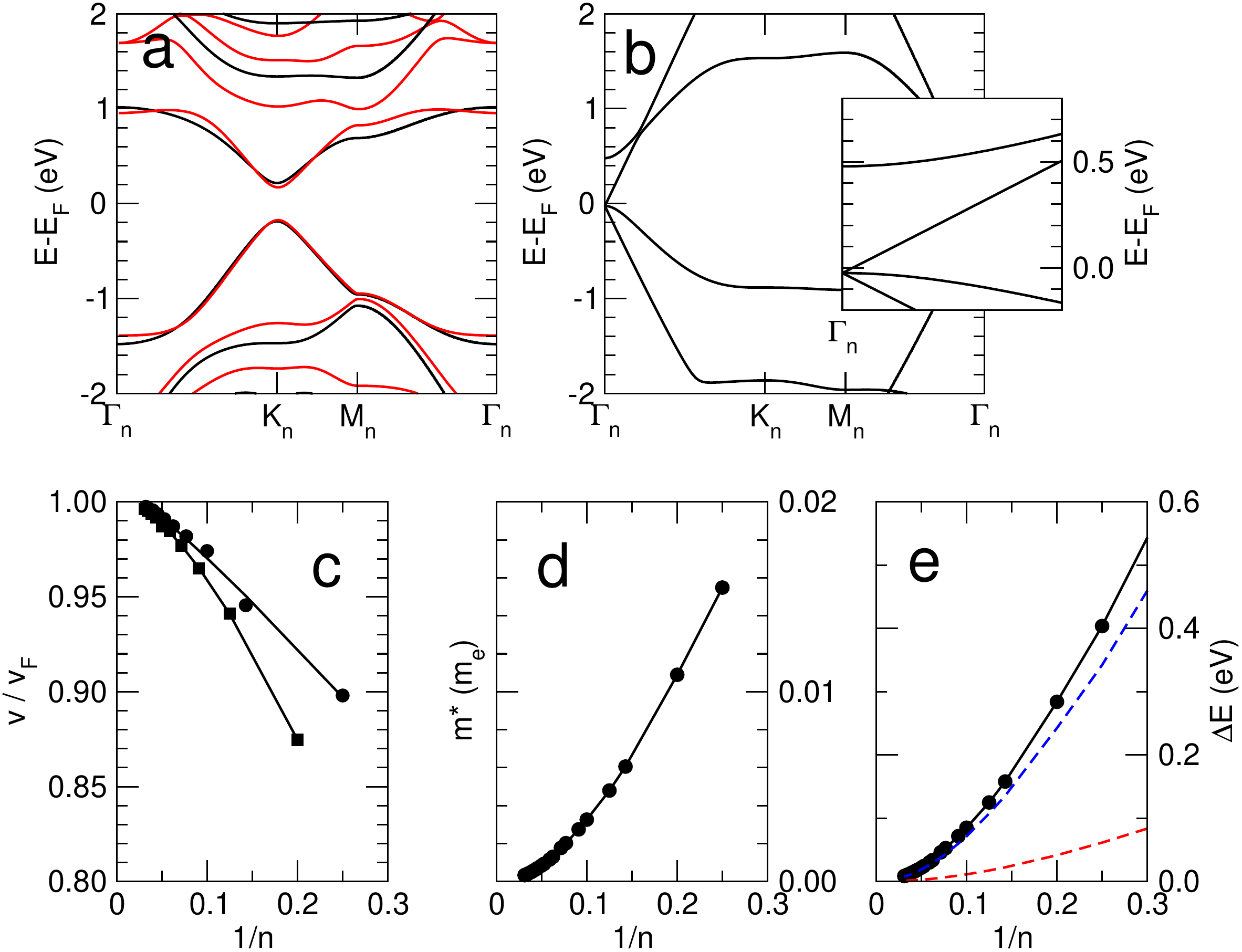}
\par\end{centering}
\caption{\label{figBN}
$a$) TB and DFT band structures for the co-doped 4x4-honeycomb.
$b$) the case of $3$x$3$ superlattice with the low-energy region in the inset. 
$c$) Effective speed, $d$) mass at the $K_n$ point versus $1/n$, and $e$) band-gap. 
The red and the blue dashed lines are the values for $n$- and $p$-doped hexagons, 
shown for comparison.}
\end{figure}
\ref{figBN} shows the computed band structure (panel (a)), together with 
the values of the effective speed of light (c), effective mass (d) and band gaps (e), 
obtained as in previous section by fitting of the numerical results,  
for different BN $n$x$n$-honeycombs. 
The results confirm the expectations, and show that such structures present a 
band-gap at the Fermi energy, compatible with pseudo-relativistic massive carriers.
Their effective rest mass is rather small, scales as $\propto{n^{-1.46}}$ 
and it is never larger than 0.016 $m_e$ for $n\ge 4$. 
This value compares favorably with the effective masses in Bi$_{1-x}$Sb$_x$ 
topological insulators ($m^*$=0.009 $m_e$) \cite{CavaBiSb}, 
and is generally lower than in bilayer graphene ($m^*$=0.03 $m_e$)\cite{Castro08} 
or in any other traditional bulk semiconductors, such as InSb ($m^*$=0.016 $m_e$). 
Since $m^*$ is the main factor affecting carrier mobility, the suggested structures 
turn out to be a good compromise between the need of opening a gap for logic applications 
and the desire of preserving the high mobility of charge carriers. 
In \ref{figBN}, panel (b) we also report the band structure of the $n=3m$ case. The structure is 
that of a zero-gap semiconductor, with two distinct charge carriers: one of them behave as electron 
(hole) in graphene, showing typical effects expected for massless carriers; 
the other is a more conventional one, with a finite excitation energy across a gap.

\section{Summary and Conclusions}\label{summary}

To summarize, we have studied the effects substitutional defects such B and N species 
have on graphene electronic structure when they are periodically arranged to form some superlattices.
Using group theoretical arguments and both TB and DFT calculations we have shown
that defects can either preserve the Dirac cones or open a band gap, 
depending on the superlattice symmetry ($D_{6h}$ and $D_{3h}$, respectively).
Specifically, honeycomb-shaped superlattices of B or N atoms give rise to $p$- and $n$-doped graphene, 
respectively, preserving the Dirac cones. 
On the other hand, when a hexagonal superlattice is formed, 
or the honeycomb one is symmetrically co-doped, the Dirac cones detach from each other to form 
a gapped, quasi-conical structure whose excitations correspond to massive Dirac particles. 
Note that this situation clearly differs form the case of randomly arranged 
B or N impurities, in which the density of states shows no band gap \cite{Roche08}.\\
For zero-gap structures the use of this superlattices offers the possibility 
to control the Fermi velocity by changing the structure periodicity, 
thereby offering the opportunity to investigate 
its role in the charge transport properties.
Differently from our recent proposal\cite{Martinazzo10},  
the gapped band structures arise because of symmetry breaking, 
as in the case of graphene interacting with a substrate such as 
SiC or BN. In the same fashion the band gap size depends 
on the superlattice periodicity.    
In our calculations we have found that gaps and charge carriers velocities
effective masses 
depends on $1/n^p$, where $p$ is in the range $1-2$, hence on some small power $0.5-1$
of the dopant concentration, and on the dopant type (B or N).
Overall the structures proposed here show a band gap larger than $k_BT$ at 
room temperature, with an effective mass generally lower that 0.01 $m_e$  
for reasonably dense meshes ($n$=4-10).
Thus, the new class of graphene structures proposed might be promising candidates for the fabrication of 
high performance interconnects, valley-based devices\cite{Xiao07}, 
but also for logic transistors, where a band gap is needed, but the extraordinary properties 
of pristine graphene need to be preserved.\\
The electronic properties of these impurities superlattices rely on symmetry, 
hence are necessarily sensitive to the dopant positions. 
As a consequence, an accurate control of the system geometry is  
necessary to exploit their properties. This might be possible in the near future 
with precise bottom-up techniques, such as the ones recently used by Ruffieux and co-workers 
\cite{RuffieuxNature, RuffieuxChemComm} to fabricate nanoribbons of well-defined widths and edges.
An indication of the possible synthetic routes together with 
formation energies for such defects superlattices can be found in the Supporting Information.
This information is available free of charge via the Internet at http://pubs.acs.org.

\section{Supporting information}

In order to compare relative structural stabilities, formation
energies for superlattice structures have been computed as follows:
\begin{equation}
\Delta H_{form} = E_{SL} - (2n^2-\eta) \mu_C - n_N \mu_N - n_B \mu_B 
\end{equation}
\begin{figure}
\begin{centering}
\includegraphics[clip,angle=-90,width=0.68\columnwidth]{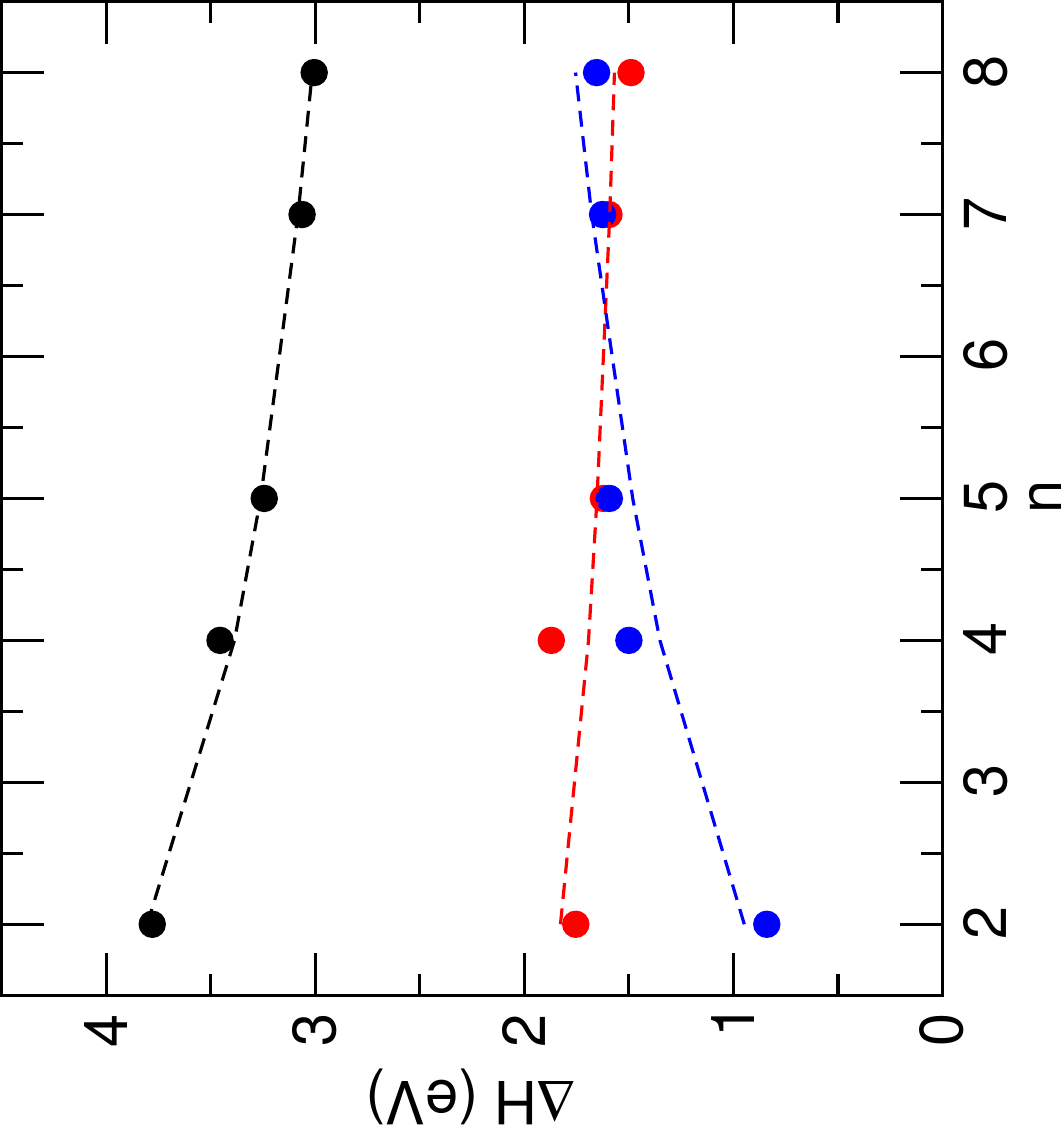}
\par\end{centering}
\caption{\label{fig:thermo}
DFT formation enthalpies (in eV) for Boron (black), Nitrogen (red) 
and co-doped (blue) honeycombs. Dashed lines show non-linear fits as a guide for the eye}
\end{figure}
where $E_{SL}$ is the total energy of the structure, $n$ is the superlattice constant,
$\eta$ is the number of dopants per cell (1 for hexagons and 2 for honeycombs), 
$n_i$ and $\mu_i$ are the number of atoms and the chemical potentials 
for each species. The chemical potentials were computed with respect to single layer graphene, 
gaseous N$_2$ and $\alpha$-Boron (in the so-called R12 structure). Density functional theory results 
are shown in \ref{tab:deltaH} and in figure \ref{fig:thermo}.\\
\begin{table}
\caption{\label{tab:deltaH} 
DFT formation enthalpies for doped superlattices for structures up to $n$=8
arranged by doping and by superlattice symmetry.
All the values shown are in eV}
\vspace{0.40cm}

\centering
\begin{tabular}{ccc|ccc}
\hline\hline
\multicolumn{3}{c}{Hexagon} & \multicolumn{3}{c}{Honeycomb}\tabularnewline
$n$ & B & N & B & N & BN\tabularnewline
\hline
2 & 1.840 & 1.078 & 3.780 & 1.754 & 0.841 \tabularnewline
4 & 1.513 & 0.878 & 3.455 & 1.871 & 1.500 \tabularnewline
5 & 1.499 & 0.933 & 3.224 & 1.622 & 1.593 \tabularnewline
7 & 1.380 & 0.676 & 3.063 & 1.596 & 1.626 \tabularnewline
8 & 1.394 & 0.695 & 3.005 & 1.490 & 2.023\tabularnewline
\end{tabular}
\end{table}

For all the cases considered the formation energy of the superlattice structures is endothermic
with respect to the pure elements with opposite trends for $n$- ($p$-)doped and co-doped structures. 
While two impurities of the same kind gain energy when lying further apart to each other, B and 
N tend instead to cluster together, in accordance with recent experimental observations \cite{ciNature10}.
However, in order to cluster the impurities have to diffuse through the graphene lattice by exchanging 
its position with a carbon atom. Such a process has a large energetic barrier being the two atoms
covalently bound, hence the kinetics of this process is expected to be extremely slow. 
We therefore suggest that these structures are expected to be stable at room temperature as well as other known structures 
(\emph{e.g.} graphene itself with respect to diamond).\\
About the practical feasibility of substitutional defects superlattices we expect they can be produced by the same
bottom-up approach  recently used to fabricate graphene structures
with atomic-scale control\cite{RuffieuxChemComm,RuffieuxNature}. 
According to this method an appropriate polyphenylene precursor, \emph{e.g.} produced by Ullmann coupling,  
undergo cyclodehydrogenation on a Cu surface to form a polycyclic aromatic hydrocarbon (PAH)\cite{Weiss99,Beernink2001,Lipton09,Treier2010}, 
that in our case should include the dopant atom. When appropriately functionalized the PAHs can then polymerize 
on the metal surface to form graphene domains up to a nanometer scale as recently shown for the synthesis of the 
atomically precise nanoribbons\cite{RuffieuxNature}.   
Such a technique has been also adopted to produce nitrogen doped fullerenes\cite{OteroFullerene} and 
two-dimensional polymers\cite{TreierJACS}, hence it 
might indeed lead to graphene-based superlattices of
substitutional defects starting from chemically doped polycyclic aromatic hydrocarbons.

\bibliography{BorazoBib}

\end{document}